\documentclass[showpacs,pra,floatfix,twocolumn]{revtex4}
\usepackage{graphicx}
\usepackage{bm,epsfig}
\usepackage{amsmath}
\usepackage{amssymb}

\begin{document}

\title{Measurement of distance and orientation of two atoms in arbitrary geometry}

\author{Qurrat-ul-Ain \surname{Gulfam}}
\email{Qurrat-ul-Ain@mpi-hd.mpg.de}
\affiliation{Max-Planck-Institut f\"ur Kernphysik, Saupfercheckweg 1, 
D-69117 Heidelberg, Germany}

\author{J\"org \surname{Evers}}
\email{joerg.evers@mpi-hd.mpg.de}
\affiliation{Max-Planck-Institut f\"ur Kernphysik, Saupfercheckweg 1, 
D-69117 Heidelberg, Germany}


\date{\today}

\begin{abstract}
Accurate measurement of relative distance and orientation of two nearby quantum particles is discussed. We are in particular interested in a realistic description requiring as little prior knowledge about the system as possible. Thus, unlike in previous studies, we consider the case of an arbitrary relative orientation of the two atoms. For this, we model the atom with complete Zeeman manifolds, and include parallel as well as orthogonal dipole-dipole couplings between all states of the two atoms. We find that it is possible to determine the distance of the two atoms independent of the orientation, as long as the particles are sufficiently close to each other. Next, we discuss how in addition the alignment of the atoms can be measured. For this, we focus on the two cases of atoms in a two-dimensional waveguide and of atoms on a surface.
\end{abstract}


\maketitle

\section{Introduction}
Progress in many areas of science and its application is fueled by the ongoing progress to measure and structure small objects. In many cases, light is used as a primary tool for reading or writing. But since light is subject to diffraction, a straightforward implementation is restricted to structures of order of the involved wavelength~\cite{rayleigh}. Different methods have been invented to surpass this limit, such as near-field imaging~\cite{near}, techniques based on the selective addressing of nearby particles~\cite{selective}, resolution enhancement due to non-classical effects~\cite{non-class}, multiphoton spectroscopy~\cite{multiphoton}, quantum lithography with classical fields~\cite{class}, or position-dependent dark states~\cite{dark}.

Among the most fundamental problems in this area is the measurement of the distance between two nearby quantum particles such as atoms. 
It has been recognized that a precise determination of the interparticle distance is possible down to distances far below the wavelength of the employed light based on their mutual interaction. For small distances, the atoms  are coupled by the dipole-dipole interaction, which modifies the optical properties of the system~\cite{Agarwal,ZF}. This was confirmed in a recent experiment~\cite{C. Hettich}, and it was found that the resonance fluorescence exhibits characteristic features which enable one to determine the relative distance over a large range of small distances~\cite{Jun-tao}. 
The resonance fluorescence has the advantage that it can be observed in the far field, and distance determination via fluorescence is not affected by the usual resolution limitations since the distance information is encoded in the frequency spectrum of the emitted light. Similar ideas have also been used for the localization of single particles~\cite{single}.

The existing distance measurement techniques based on the dipole-dipole interaction, however, are restricted to two two-level atoms in specific geometries, such as aligned along the propagation direction of the exciting laser field. In most practical cases, however, the relative orientation of the two nearby atoms is unknown, and for similar reasons, it is equally difficult to measure the relative orientation as the distance. Thus the question arises, whether the ideas of~\cite{C. Hettich,Jun-tao} can be extended to the case of arbitrary orientation. It turns out that it is not meaningful to study the system of two two-level atoms in the case of arbitrary orientations. The reason for this is the appearance of dipole-dipole couplings between orthogonal transition dipole moments (DDOTDM) in more general geometries~\cite{Agarwal2001,MK,MKJ,JE,Sandra}. The electric field emitted by one of the particles has not only a component corresponding to the emitting transition dipole moment, but also a component along the interparticle distance vector. The projection of the latter field component on a transition dipole moment in the second atom can be non-zero even if it is orthogonal to the emitting dipole~\cite{JE}. In a real atom with magnetic level structure, these DDOTDM  lead to the population of excited states even if they are not driven by the external laser field. Thus, the two-level approximation breaks down, and correct predictions can only be expected if the theoretical modelling includes complete Zeeman manifolds including all occurring dipole dipole couplings~\cite{MK}.

Motivated by this, here we study the determination of relative distance and orientation of two nearby atoms in arbitrary geometry. The atoms are driven by a single resonant standing-wave laser field, and we make use of the far field resonance fluorescence intensity and spectrum as observables.  Each atom is modelled as a four-level system with one ground state (total angular momentum zero) and three excited states (total angular momentum one), including all relevant dipole-dipole couplings occurring in arbitrary geometries. We start by analyzing the distance determination for the case of a known orientation, and present dressed-state interpretations of the obtained  resonance fluorescence spectra in various cases of relevance. Next, we describe a method to determine the interparticle distance for arbitrary orientation, which works as long as the particles are sufficiently close to each other. Finally, we discuss methods to determine the relative orientation of the two particles, focusing  on the two cases of atoms confined in a planar waveguide and atoms on a surface. 

\begin{figure}
\includegraphics[width=0.9\linewidth ]{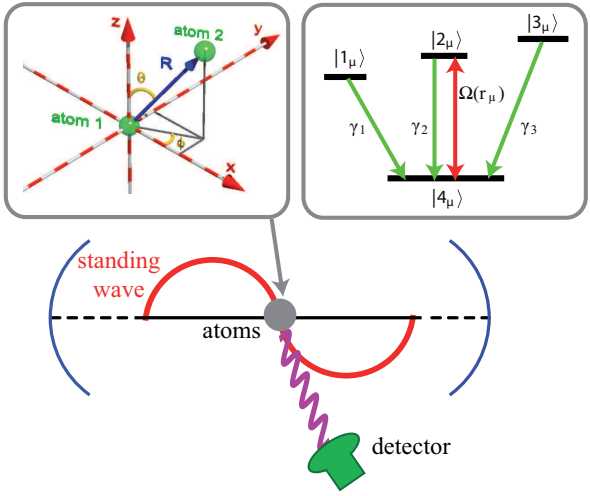}
\caption{\label{system}(Color online) Scheme for the determination of relative distance and orientation of two nearby atoms. The atoms $\mu\in\{1,2\}$ are driven on one transition by a standing wave laser field with Rabi frequency $\Omega({\bf r}_\mu)$ and scatter light, which is registered in the far field with a detector. The interatomic distance vector $\textbf{\textit{R}}$ is arbitrary, as shown in the left inset. The right inset shows the level structure of each atom. Each atom has a single ground state (zero angular momentum) and three excited states (angular momentum 1).}
\end{figure}

\section{Theoretical considerations}

\subsection{Master equation}

Our system consists of two identical nearby  atoms located at
$\textbf{r}_{i}$ ($i\in\{1,2\}$) as shown in Fig.~\ref{system}. Each atom is modelled as a four-level system with one ground and three excited states, modelling the complete magnetic substructure of a transition from an $S$ state (angular momentum 0) to a $P$ state (angular momentum 1). The Hamiltonian of the two atoms coupled to the surrounding vacuum field is given by
\begin{equation}
H = H_A+H_F+H_{vac}\,,
\end{equation}
where
\begin{subequations}
\begin{align}
H_A&=\hbar \sum_{\mu=1}^{2}\sum_{i=1}^{3}\omega_iS_{i+}^{(\mu)}S_{i-}^{(\mu)}\,,\\
H_F&=\sum_{\textbf{\textit{k}}s}\hbar\omega_ka_{\textbf{\textit{k}}s}^{\dag}a_{\textbf{\textit{k}}s}\,,\\
H_{vac}&=-\hat{\textbf{\textit{d}}}^{(1)}\cdot\hat{\textbf{\textit{E}}}(\textbf{\textit{r}}_1)-\hat{\textbf{\textit{d}}}^{(2)}\cdot\hat{\textbf{\textit{E}}}(\textbf{\textit{r}}_2)\,.
\end{align}
\end{subequations}
In these equations, $H_A$ is the Hamiltonian that describes the free evolution of the bi-atomic system. We set the energy of the ground state to zero, and the energies of the excited atomic states $|i\rangle$ ($i\in\{1,2,3\}$) are $\hbar\omega_i$. The raising and lowering operators on the $| 4_\mu\rangle$ $\leftrightarrow | i_\mu\rangle$ transition of atom $\mu$ $(\mu\in{1,2})$ are 
\begin{equation}
S_{i+}^{(\mu)}=| i_\mu\rangle\langle4_\mu| \quad {\rm{and}}  \quad S_{i-}^{(\mu)}=| 4_\mu\rangle\langle i_\mu|.
\end{equation}
The Hamiltonian of the vacuum field is described by $H_F$, with $a_{\textbf{\textit{k}}s}$ and $a_{\textbf{\textit{k}}s}^\dag$ as the field annihilation and creation operators.
$H_{vac}$ describes the interaction of the atom with the vacuum field in the dipole approximation with the vacuum field $\hat{\textbf{\textit{E}}}(\textbf{r})$ given by
\begin{equation}
\hat{\textbf{\textbf{E}}}(\textbf{r})=\iota\sum_{\textbf{\textit{k}}s}\sqrt{\frac{\hbar\omega_k}{2\epsilon_0V}}\epsilon_{\textbf{\textit{k}}s}e^{\iota \textbf{\textit{k}}\cdot \textbf{\textbf{r}}}a_{\textbf{\textit{k}}s}+\rm{H\ldotp c\ldotp}\,.
\end{equation}
 $\textbf{\textit{k}}$ is the wave vector, $\epsilon_{\textbf{\textit{k}}s}$ the polarization, $\omega_k$ the frequency of a field mode,  and $V$ the quantization volume. 
We use the Wigner-Eckart theorem~\cite{Sakurai} to determine the electric dipole moment operator of atom $\mu$ which is given by 
\begin{equation}
 \hat{\textbf{\textit{d}}}^{(\mu)}=\sum_{i=1}^{3}\textbf{\textit{d}}_{i}S_{i+}^{(\mu)}+\rm{H.c.}\,,
\end{equation}
and the dipole moments $\textbf{\textit{d}}_{i}=\langle i|\hat{\textbf{\textit{d}}}|4\rangle$ are given by the matrix elements of the electric dipole moment operator $\hat{\textbf{\textit{d}}}$ as
\begin{equation}
\textbf{\textit{d}}_{1}=\mathcal{D} \epsilon^{(+)}, \quad \textbf{\textit{d}}_{2}=\mathcal{D}\textbf{\textit{e}}_{z}, \quad \textbf{\textit{d}}_{3}=-\mathcal{D}\epsilon^{(-)}\,.
\end{equation}
$\epsilon^{(\pm)}=(\textbf{\textit{e}}_x \pm\iota \textbf{\textit{e}}_{y})/\sqrt{2}$, and $\mathcal{D}$ denotes the reduced dipole matrix element.
Note that the dipole moments $\textbf{\textit{d}}_{i}$ are independent of the atomic index $\mu$ since the two atoms are identical.
The vector that defines the relative position of the atoms in spherical coordinates is given by
\begin{equation}
\textbf{\textit{R}}=\textbf{r}_{2}-\textbf{r}_{1}=R(\sin\theta\cos\phi,\sin\theta\sin\phi,\cos\theta)^T\,.
\end{equation}

\begin{figure}[t]
\includegraphics*[width=0.9\linewidth,angle=0]{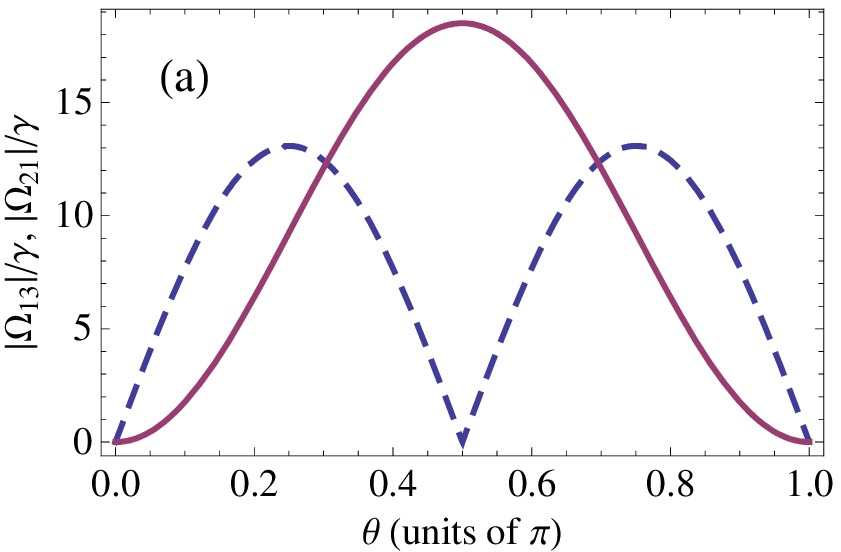}
\includegraphics*[width=0.9\linewidth,angle=0]{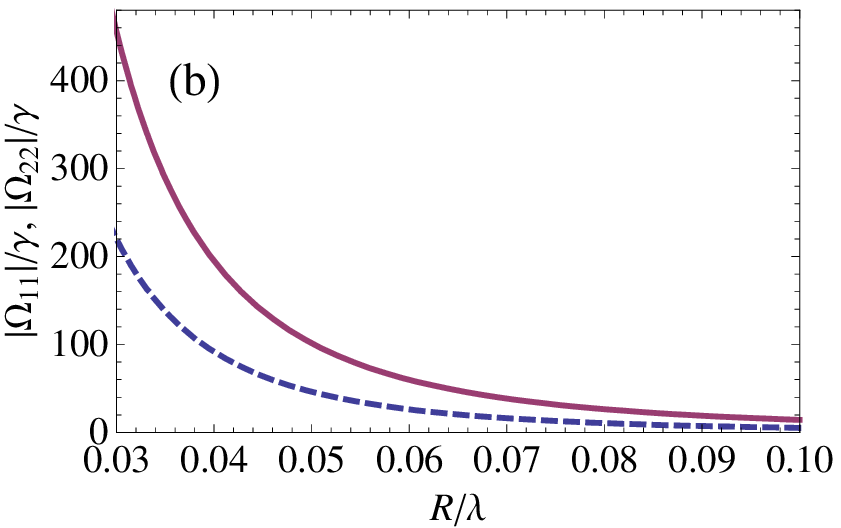}

\caption{\label{fig-coupling}(Color online) (a) Magnitude of few orthogonal dipole-dipole coupling constants. The parameters are $R=0.08 \lambda$ and $\phi=\pi/2$. The red solid curve shows $|\Omega_{13}|$, while the blue dashed curve depicts $|\Omega_{21}|=|\Omega_{32}|$. (b) Magnitude of few parallel dipole-dipole coupling constants at $\theta=\pi$. The red solid line shows $|\Omega_{22}| $, and the dashed blue line depicts $|\Omega_{11}|=|\Omega_{33}|$.}
\end{figure}

Using standard methods, the system Master equation evaluates to~\cite{Agarwal2001,JE,MKJ,MK}
\begin{equation}\label{master}
\dfrac{\partial \varrho}{\partial t}=-\dfrac{\iota}{\hbar}[H_{A},\varrho]-\dfrac{\iota}{\hbar}[H_{\Omega},\varrho]+\mathcal{L}_{\gamma}\varrho.
\end{equation}
The Hamiltonian $H_{\Omega}$ arises from the coherent part of the dipole-dipole interaction and is given by
\begin{align}
H_{\Omega}&=-\hbar\sum_{i=1}^{3}\{\Omega_{ii}S_{i+}^{(2)}S_{i-}^{(1)}+\rm{H.c.}\}-\hbar\{\Omega_{21}(S_{2+}^{(2)}S_{1-}^{(1)}\nonumber\\
&+S_{2+}^{(1)}S_{1-}^{(2)})+\rm{H.c.}\}-\hbar\{\Omega_{31}(S_{3+}^{(2)}S_{1-}^{(1)}+S_{3+}^{(1)}S_{1-}^{(2)})\nonumber\\
&+\rm{H.c.}\}-\hbar\{\Omega_{32}(S_{3+}^{(2)}S_{2-}^{(1)}+S_{3+}^{(1)}S_{2-}^{(2)})+\rm{H.c.}\}\,.
\end{align}
The incoherent part of Eq.~(\ref{master}) is given by
\begin{align}
\mathcal{L}_{\gamma}\varrho=&-\sum_{\mu=1}^{2}\sum_{i=1}^{3}\gamma_{i}(S_{i+}^{(\mu)}S_{i-}^{(\mu)}\varrho+\varrho S_{i+}^{(\mu)}S_{i-}^{(\mu)}-2S_{i-}^{(\mu)}\varrho S_{i+}^{(\mu)})\nonumber\\
&-\sum_{i=1}^{3}\{\Gamma_{ii}(S_{i+}^{(2)}S_{i-}^{(1)}\varrho+\varrho S_{i+}^{(2)}S_{i-}^{(1)}-2S_{i-}^{(1)}\varrho 
S_{i+}^{(2)})+\rm{H.c.}\}\nonumber\\
&-\sum^{2}_{\substack {\mu,\nu=1\\ \mu\neq\nu}}\{\Gamma_{21}(S_{2+}^{(\mu)}S_{1-}^{(\nu)}\varrho+\varrho 
S_{2+}^{(\mu)}S_{1-}^{(\nu)}-2S_{1-}^{(\nu)}\varrho S_{2+}^{(\mu)})\nonumber\\
&+\Gamma_{31}(S_{3+}^{(\mu)}S_{1-}^{(\nu)}\varrho+\varrho 
S_{3+}^{(\mu)}S_{1-}^{(\nu)}-2S_{1-}^{(\nu)}\varrho S_{3+}^{(\mu)})\nonumber\\
& +\Gamma_{32}(S_{3+}^{(\mu)} S_{2-}^{(\nu)}\varrho+\varrho S_{3+}^{(\mu)}S_{2-}^{(\nu)}-2S_{2-}^{(\nu)}\varrho S_{3+}^{(\mu)})+ \rm{H.c.}\}\,.
\end{align}
In the above equations, coefficients $\Omega_{ij}$ and $\Gamma_{ij}$ with $i=j$ represent dipole-dipole couplings between a transition dipole of one atom and the corresponding parallel dipole of the other atom, whereas terms with $i\neq j$ correspond to dipole-dipole couplings between orthogonal transition dipole moments in the two atoms.
The coefficients $\Omega_{ij}$ and $\Gamma_{ij}$ can be calculated as
\begin{subequations}
\begin{align}
\Omega_{ij}=\dfrac{1}{\hbar}[\textbf{\textit{d}}_{i}^{T} \textit{\rm{Re}}(\overleftrightarrow{\mathcal{X}})\textbf{\textit{d}}_{j}^{\ast}]\,, \label{oddes}\\
\Gamma_{ij}=\dfrac{1}{\hbar}[\textbf{\textit{d}}_{i}^{T} \textit{\rm{Im}}(\overleftrightarrow{\mathcal{X}})\textbf{\textit{d}}_{j}^{\ast}]\,.\label{oddgam}
\end{align}
\end{subequations}
Here, $\rm{Re}(\overleftrightarrow{\mathcal{X}})$ [$\rm{Im}(\overleftrightarrow{\mathcal{X}})$] represents the real [imaginary] part of the tensor $\overleftrightarrow{\mathcal{X}}$ whose components $\overleftrightarrow{\mathcal{X}_{kl}}$ ($k,l\in \{1,2,3\}$) are given by
\begin{align}
\overleftrightarrow{\mathcal{X}_{kl}}(\textbf{\textit{R}})&=\dfrac{k_{o}^{3}}{4 \pi \varepsilon_{0}}[\delta_{kl}(\dfrac{1}{\eta}+\dfrac{\iota}{\eta^{2}}-\dfrac{1}{\eta^{3}})\nonumber\\
&-\dfrac{\textbf{\textit{R}}_{k}\textbf{\textit{R}}_{l}}{R^{2}}(\dfrac{1}{\eta}+\dfrac{3\iota}{\eta^{2}}-\dfrac{3}{\eta^{3}})]e^{\iota\eta}.
\end{align}
$\delta_{kl}$ is the Kronecker delta symbol and $\eta=k_{0}R$, with the approximation $\omega_{1}\approx\omega_{2}\approx\omega_{3}\approx\omega_{0}$, where $\omega_{0}=ck_{0}$ is the mean transition frequency.
From Eq.~(\ref{oddes}), the different dipole-dipole coupling constants $\Omega_{ij}$ evaluate to
\begin{subequations}
\begin{align}
 \Omega_{31}&=\gamma\dfrac{3}{4\eta^{3}}[(\eta^{2}-3)\cos\eta-3\eta\sin\eta]\sin^{2}\theta e^{-2\iota\phi}\,,\\
 \Omega_{11}&=3\dfrac{\gamma}{8\eta^{3}}\{[3\eta^{2}-1+(\eta^{2}-3)\cos2\theta]\cos\eta\nonumber\\
&-\eta(1+3\cos2\theta)\sin\eta\}\,, \label{om11}\\
 \Omega_{21}&=-\sqrt{2}\cot\theta\Omega_{31}e^{\iota\phi}\,,\\
\Omega_{22}&=\Omega_{11}-(2 \cot^{2}\theta-1)\Omega_{31}e^{2\iota\phi}\,, \label{om22}\\
 \Omega_{32}&=-\Omega_{21}, \quad \Omega_{33}=\Omega_{11}\,.
\end{align}
\end{subequations}
The corresponding incoherent coupling constants $\Gamma_{ij}$ follow from Eq.~(\ref{oddgam}) as
\begin{subequations}
\begin{align}
 \Gamma_{31}&=\gamma\dfrac{3}{4\eta^{3}}[(\eta^{2}-3)\sin\eta+3\eta\cos\eta]\sin^{2}\theta e^{-2\iota\phi}\,,\\
 \Gamma_{11}&=3\dfrac{\gamma}{8\eta^{3}}\{[3\eta^{2}-1+(\eta^{2}-3)\cos2\theta]\sin\eta\nonumber\\
&+\eta(1+3\cos2\theta)\cos\eta\}\,,\\
 \Gamma_{21}&=-\sqrt{2}\cot\theta\Gamma_{31}e^{\iota\phi}\,,\\
 \Gamma_{22}&=\Gamma_{11}-(2 \cot^{2}\theta-1)\Gamma_{31}e^{2\iota\phi}\,,\\
 \Gamma_{32}&=-\Gamma_{21}, \quad \Gamma_{33}=\Gamma_{11}\,.
\end{align}
\end{subequations}
Some examples illustrating the dependence of the coupling constants $\Omega_{ij}$ on $\theta$ or $R$ are shown in Fig.~\ref{fig-coupling}.

The total spontaneous decay rate of each individual atom is given by $2\gamma_{i}$, where
\begin{equation}
 \gamma := \gamma_{i}=\dfrac{1}{4\pi \varepsilon_{0}}\dfrac{2|\textbf{\textit{d}}_{i}|^{2}\omega_{0}^{3}}{3\hbar c^{3}}\,,
\end{equation}
and we have again used the approximation $\omega_{i}\approx\omega_{0}$. 

%

We finally include an external driving laser field, which we assume to be polarized along the z-axis. Its electric field is given by 
\begin{equation}
\textbf{\textit{E}}_L=\mathcal{E}_z \textbf{\textit{e}}_{z} e^{\iota \textbf{\textit{k}}_{L}\cdot \textbf{r}} e^{-\iota\omega_{L}t}+\rm{c.c.}\,,
\end{equation}
where $\mathcal{E}_z$ denotes the amplitude, $\omega_{L}$  the frequency, and $\textbf{\textit{e}}_{z}$ is the polarization of the field, and $\rm{c.c.}$ denotes the complex conjugate. The wave vector $\textbf{\textit{k}}_{L}$ with wave number $\textit{k}_{L}=2 \pi/\lambda$ points along the positive x-axis. 

In a suitable interaction picture, we obtain
\begin{equation}
\dfrac{\partial \tilde{\varrho}}{\partial t}=-\dfrac{\iota}{\hbar}[\tilde{H}_L+\tilde{H}_{A},\tilde{\varrho}]-\dfrac{\iota}{\hbar}[H_{\Omega},\tilde{\varrho}]+\mathcal{L}_{\gamma}\tilde{\varrho}\,,
\end{equation}
with
\begin{equation}
\label{ham-laser}
 \tilde{H}_{A}=-\hbar \sum_{\mu=1}^{2}\sum_{i=1}^{3}\Delta_iS_{i+}^{(\mu)}S_{i-}^{(\mu)}\,.
\end{equation} 
Here, the detunings $\Delta_{i}=\omega_{L}-\omega_{i}\quad (i\in\{1,2,3\})$.
The interaction of the system with the external laser field in the electric dipole and the rotating wave approximation is described by
\begin{equation}
\tilde{H}_L = -\frac{\hbar}{2}\sum_{\mu=1}^2[\Omega(\textbf{\textit{r}}_{\mu})S_{2+}^{(\mu)}+\rm{H\ldotp c\ldotp]}\,.
\end{equation}
The position dependent Rabi frequencies are given by
\begin{align}
\label{rabifreq}
 \Omega(\textbf{r}_\mu)=\Omega\sin(\textbf{\textit{k}}_{L}\cdot \textbf{r}_\mu)\,,
\end{align}
where $\Omega=\mathcal{D}\mathcal{E}_z/\hbar$ ($\mu\in\{1,2\}$).  Due to its polarization, the laser field couples only to the $|2\rangle\leftrightarrow|4\rangle$ transitions in the two atoms.

\subsection{Observables}

Our observables are the resonance fluorescence intensity and the resonance fluorescence spectrum of the light emitted by the two atoms. The spectrum of resonance fluorescence up to a geometrical factor is determined by real part of the Fourier transform of the two time correlation function of the electric field~\cite{spectrum},
\begin{equation}
S(\omega)={\rm{Re}}\int^{\infty}_{-\infty}e^{-\iota \omega \tau}\langle\hat{\textbf{\textit{E}}}^{(-)}(\textbf{r},t+\tau) \cdot \hat{\textbf{\textit{E}}}^{(+)}(\textbf{r},t)\rangle_{st}d\tau\,. \label{spectrum}
\end{equation}
In this equation, $\hat{\textbf{\textit{E}}}^{(-)}$ $[\hat{\textbf{\textit{E}}}^{(+)}]$ denotes the positive [negative] frequency part of the electric field operator and the subscript $st$ refers to the steady state. At a point $\textbf{r}=r\hat{\textbf{r}}$ in the far-field zone, the negative frequency part of the electric field operator evaluates to~\cite{Agarwal}
\begin{align}
\hat{\textbf{\textit{E}}}^{(-)}&(\textbf{r},t)=\hat{\textbf{\textit{E}}}_{free}^{(-)}(\textbf{r},t) \nonumber\\
&-\dfrac{1}{4\pi\epsilon_{0}c^{2}r}\sum_{i=1}^{4}\omega^{2}_{i}\:\hat{\textbf{r}}\times(\hat{\textbf{r}}\times \textbf{\textit{d}}_{i})\:\widetilde{S}_{i+}(\hat{t})e^{\iota \omega_{L}\hat{t}}, \label{negfield}
\end{align}
where $\hat{t}=t-r/c$ is the retarded time and $\widetilde{S}_{i\pm}(t)=\exp(\mp\iota \omega_{L}t)S_{i\pm}(t)$. The first term denoting the free field can be neglected if the point of observation lies outside the driving field.

The resonance fluorescence intensity is given by the one-time normally ordered correlation function of the electric field operators,
\begin{equation}
\textbf{I}_{st}=\langle\hat{\textbf{\textit{E}}}^{(-)}(\textbf{r},t) \cdot \hat{\textbf{\textit{E}}}^{(+)}(\textbf{r},t)\rangle_{st}\,.
\end{equation}
In the following, the point of observation is assumed to be along the $y$ or $-x$ direction for the resonance fluorescence spectrum, and along the $z$ direction for the resonance fluorescence intensity. Evaluating the cross products in Eq.~(\ref{negfield}), we find that our choice of observation direction enables us to separate linearly polarized light emitted on transitions $|2\rangle \leftrightarrow |4\rangle$ from the circularly polarized light emitted on transitions $|i\rangle \leftrightarrow |4\rangle$ ($i\in\{1,3\}$) by means of a polarization analyzer. We designate linearly [circularly] polarized spectra as  $\pi$ [$\sigma$] ones, respectively.

\section{RESULTS}

The resonance fluorescence spectrum Eq.~(\ref{spectrum}) emitted by the two atoms in general is rather complicated, but it simplifies considerably in certain parameter cases, as it was found already in the case of two nearby two-level systems~\cite{Jun-tao}. In the following, we will in particular refer to the case of either small or large interatomic spacing, on a length scale given by the involved transition wavelength. For small distance, the coherent part of the dipole-dipole interaction dominates the system dynamics, with corrections due to the much weaker laser field Rabi frequencies. In the opposite case of larger separation, the Rabi frequencies dominate, with corrections from the dipole-dipole interaction. Spectra for situations in which the dipole-dipole interaction and the Rabi frequencies are comparable usually can not be interpreted in a straightforward way. In these cases, the driving field intensity can be increased or decreased in order to evolve in one of the two simpler cases.
In the following, we will make use of this general observation, and present our results in two steps. First, we will describe methods to determine the interatomic separation in various cases of interest. Second, we will discuss the determination of the relative orientation of the two atoms.

For the numerical analysis, we assume that $\textbf{r}_{1}=(0.05\lambda,0,0)$. Our measurement techniques, however, also apply to other values of $\textbf{r}_{1}$. A special case arises if one of the atoms is at a node of the standing wave field. Such situations can be circumvented by shifting the phase of the standing wave slightly.
We also assume the resonance condition, \textit{i.e.}, $\Delta_i=0$, see Eq.~(\ref{ham-laser}).

\subsection{Determination of the interatomic separation}

\begin{figure}[t]
\includegraphics*[width=0.9\linewidth,angle=0]{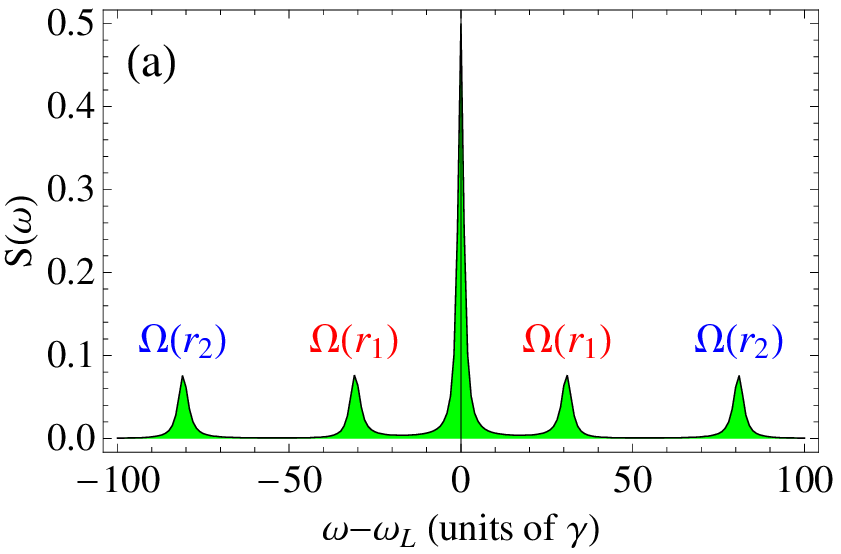}
\includegraphics*[width=\linewidth,angle=0]{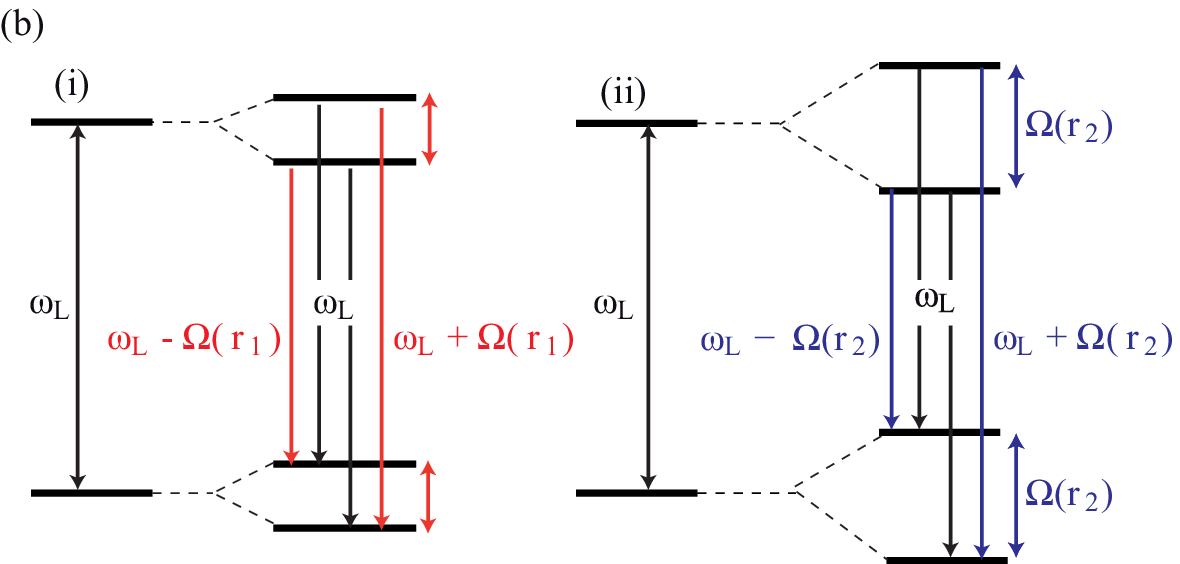}
\caption{\label{fig-large}(Color online) (a) Incoherent spectrum of resonance fluorescence for the case of larger atomic distances. The parameters are $R=0.3 \lambda$, $\phi=0$, $\theta=\pi/2$ and $\Omega=100\gamma$. The side peaks are located at the Rabi frequencies, $\Omega(\textbf{r}_{1})\approx30.90\gamma$ and $\Omega(\textbf{r}_{2})\approx80.90\gamma$.
(b) The dressed state representation of the system in (a) showing the ac-Stark splitting of the two atoms. The splitting in each atom corresponds to the respective position dependent Rabi frequency.
}

\end{figure} 

\subsubsection{\label{fixed}Known orientation of the atoms}

As a first step, we will present results for the situation of a fixed, known orientation. In particular, for simplicity,  we analyze the case $\theta = \pi/2, \phi = 0$. Then, the orthogonal dipole-dipole couplings $\Omega_{2i}$ and $\Gamma_{2i}$ ($i\in\{1,3\}$) vanish. Thus, the population is trapped only in the levels $|2_\mu\rangle$ and $|4_\mu\rangle$, and the system essentially reduces to that of two two-level atoms~\cite{Jun-tao}. This simple case, however, suffices for the main aim of this section, which is to extend previous results by interpreting each case in detail in terms of the corresponding dressed state picture.  This insight will enable us to explain our later main results for general geometries.
In principle, the results in this section also generalize to more complicated known orientations. While then the dressed-state analysis is in complete analogy to our discussions here, the analytical expressions are considerably more complicated. In any case, if the orientation is known, a numerical fit of the measured spectrum leads to the desired distance information. In the following Section~\ref{unknown}, we will extend our analysis to arbitrary orientations.

\begin{figure}[t]
\includegraphics*[width=0.9\linewidth,angle=0]{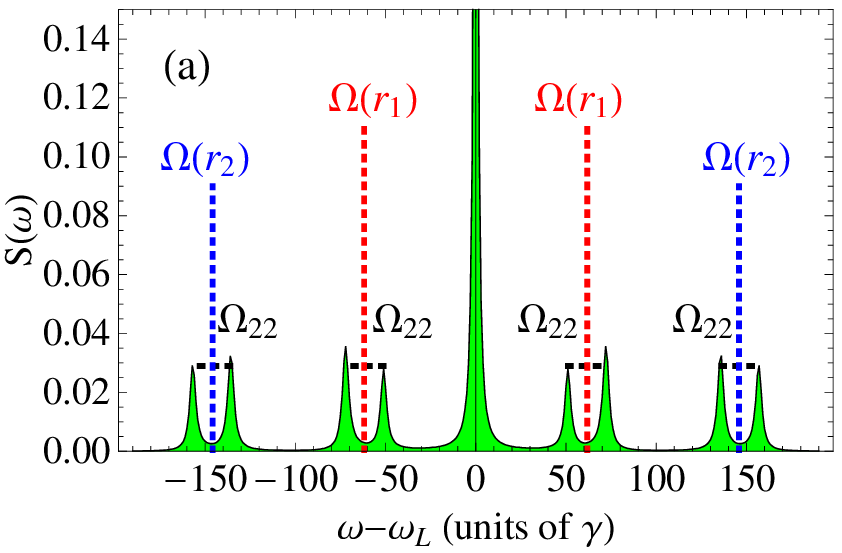}
\includegraphics*[width=\linewidth,angle=0]{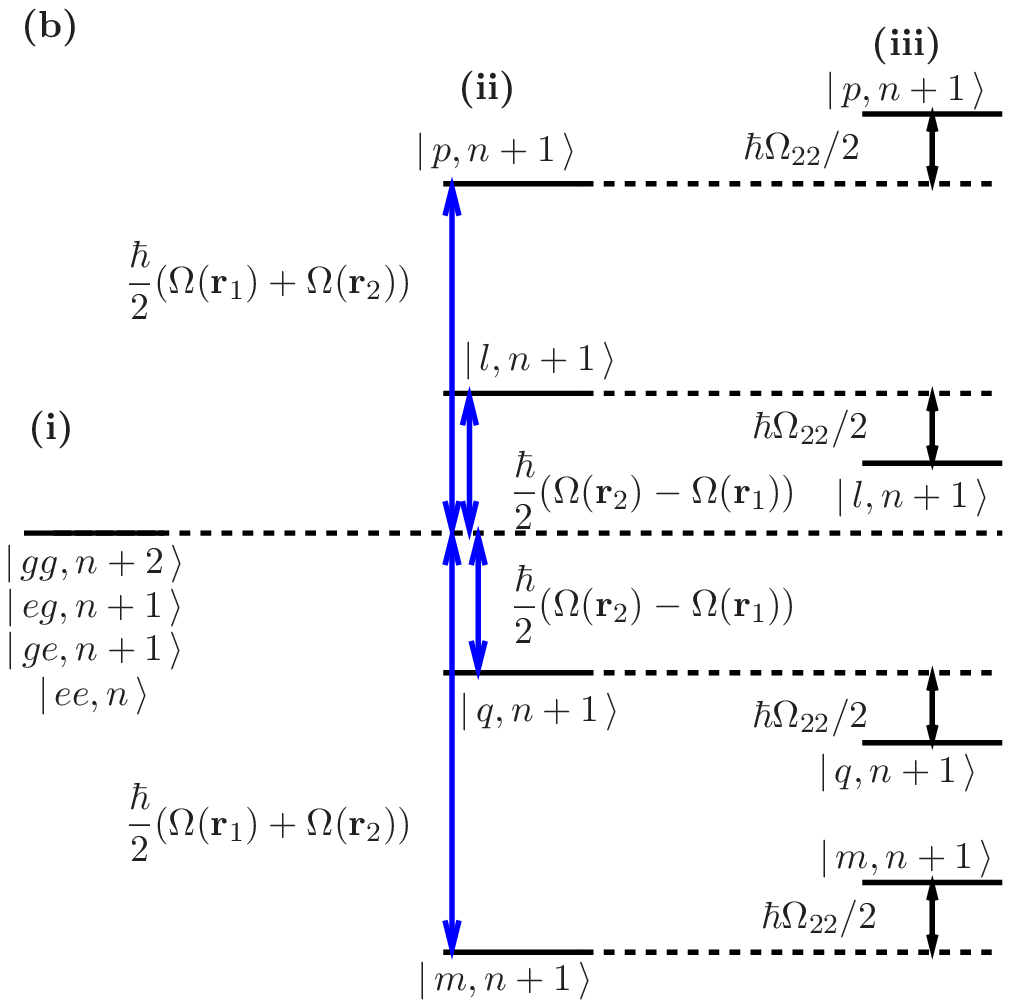}
\caption{\label{fig-middle} (Color online) (a) Incoherent spectrum of resonance fluorescence for intermediate distances. The parameters are $R=0.08\lambda$, $\theta=\pi/2$, $\phi=0$ and $\Omega=200\gamma$, such that the laser Rabi frequencies are larger than the relevant dipole-dipole coupling constants. The centre of the side band structures are located at the respective position dependent Rabi frequencies, $\pm \Omega(\textbf{r}_{1})\approx \pm 61.8\gamma$ and $\pm  \Omega(\textbf{r}_{2})\approx \pm 145.79\lambda$.
(b) Dressed-state representation. (i) shows the system without couplings. (ii) includes the dominant laser field couplings. (iii) in addition includes corrections due to the dipole-dipole interaction. 
}
\end{figure} 


For larger interatomic separation (about $\lambda/10\leq R \leq\lambda/2$), the dipole dipole interaction is almost negligible and the two atoms can be considered independent. A typical spectrum is shown in Fig.~\ref{fig-large}(a), and consists of five peaks. The middle peak occurs at $\omega_L$, and the two side peak doublets are symmetrically located at $\omega_L\pm\Omega(\textbf{\textit{r}}_{\mu})$. An interpretation of the peak structure in terms of the system dressed states is shown in Fig.~\ref{fig-large}(b). It turns out that the sidebands are shifted by the position-dependent Rabi frequencies experienced by the two atoms, respectively. Thus, the peak positions give a direct measure of the position of each atom and hence the interatomic separation can be calculated using the expressions for the position-dependent Rabi frequencies Eq.~(\ref{rabifreq})~\cite{Jun-tao}. The atomic levels in atom $\mu$ split by the amount $\Omega(\textbf{\textit{r}}_{\mu})$, just as in the Mollow spectrum~\cite{Scully}.
Note that the two split doublets may coincide, for example, for $\theta\in\{0,\pi\}$, since then $\Omega(\textbf{\textit{r}}_{1}) = \Omega(\textbf{\textit{r}}_{2})$.


For intermediate interatomic distances (approx. $\lambda/30\leq R\leq\lambda/10$), the dipole dipole interaction becomes relevant. If the parameters are such that the driving field Rabi frequency is comparable to the dipole-dipole interaction, then the spectrum obtained is difficult to interpret. This, for example, occurs if the distance is decreased to $R=0.08\lambda$ in Fig.~\ref{fig-large}(a). We thus increase the Rabi frequency of the driving laser field and obtain the spectrum shown in Fig.~\ref{fig-middle}(a). It can be observed that the sidebands are split by the dipole-dipole interaction. The corresponding dressed-state picture is shown in Fig.~\ref{fig-middle}(b). 
This picture is obtained by evaluating the eigenstates and eigenvalues of the interaction picture Hamiltonian in the limit $\Omega(\textbf{r}_{i})\gg\Omega_{22}$. 
We consider the case $\Omega(\textbf{r}_{1})\neq\Omega(\textbf{r}_{2})$ and assume  $\Omega(\textbf{r}_{2})>\Omega(\textbf{r}_{1})$. In part (i) of Fig.~\ref{fig-middle}(b), there is no interaction between the two atoms, such that four atom-field states are degenerate. In (ii), we include the interaction with the strong driving field, which splits the level scheme into four dressed states. The corresponding dressed states are listed in Tab.~\ref{my}. Finally, in part (iii), the weak dipole-dipole interaction is added, which further shifts the eigenenergies of all dressed states by approximately $\pm\Omega_{22}/2$.

\begin{figure}[t]
\includegraphics[width=0.9\linewidth ]{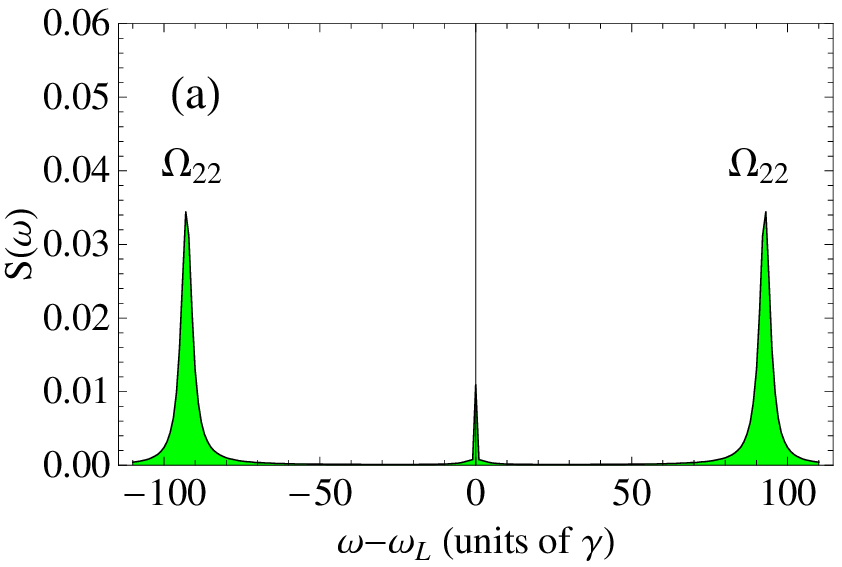}
\includegraphics*[width=0.7\linewidth]{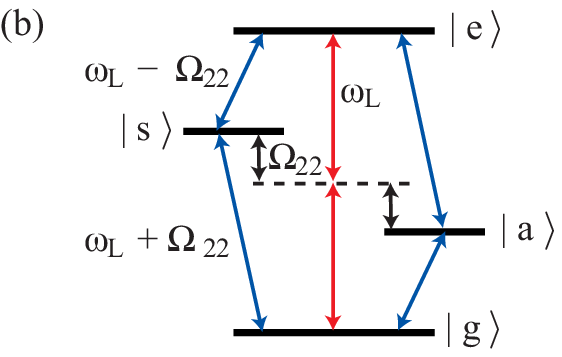}
\caption{\label{fig-small} (Color online) Incoherent spectrum of resonance fluorescence in the small distance case, with laser field Rabi frequencies negligible compared to the dipole-dipole coupling constants. The parameters are $R=0.04 \lambda$, $\theta=\pi/2$, $\phi=0$ and $\Omega=20\gamma$. The side peaks occur at $\omega_L\pm\Omega_{22}$.
(b) Dressed state representation. The energies of the symmetric $|s\rangle$ and the antisymmetric $|a\rangle$ states are shifted by the dipole dipole coupling parameter $\Omega_{22}$. 
}
\end{figure}

\begin{table}[t]
\begin{center}
\begin{tabular}{ l l l  }
\hline \hline
  Dressed state & Composition & Energy \\
	\hline
  $|p,n+1\rangle$ & $(1,1,1,1)/2$ & $\hbar(\Omega(\textbf{r}_{1})+\Omega(\textbf{r}_{2})+\Omega_{22})/2$\\
  $|m,n+1\rangle$ & $(1,-1,-1,1)/2$ & $-\hbar(\Omega(\textbf{r}_{1})+\Omega(\textbf{r}_{2})-\Omega_{22})/2$\\
  $|q,n+1\rangle$ & $(-1,-1,1,1)/2$ & $\hbar(\Omega(\textbf{r}_{1})-\Omega(\textbf{r}_{2})-\Omega_{22})/2$\\
  $|l,n+1\rangle$ & $(-1,1,-1,1)/2$ & $-\hbar(\Omega(\textbf{r}_{1})-\Omega(\textbf{r}_{2})+\Omega_{22})/2$\\
\hline \hline
\end{tabular}
\end{center}
\caption{\label{my}Eigenvectors and eigenvalues of the interaction Hamiltonian of two two-level atoms in the limit $\Omega(\textbf{r}_{i}) \gg \Omega_{22}$.}
\end{table}

There are sixteen possible transitions between the eigenkets having $n+1$ and $n$ photons. Transitions $|p,n+1\rangle\rightarrow|p,n\rangle$, $|l,n+1\rangle\rightarrow|l,n\rangle$ and $|q,n+1\rangle\rightarrow|q,n\rangle$ and $|m,n+1\rangle\rightarrow|m,n\rangle$ correspond to transition frequency $\omega_{L}$. The eight transitions having frequencies $\omega_{L}\pm(\Omega(\textbf{r}_{i})\pm\Omega_{22})$, $i\in(1,2)$ form the two prominent side band doublets on each side (eight side peaks in total) visible in Fig.~\ref{fig-middle}(a). These peaks are crucial for the distance determination. Setting parameters such that  $\Omega(\textbf{r}_{1})=\Omega(\textbf{r}_{2})$ would reduce the number of transitions $\omega_{L}\pm(\Omega(\textbf{r}_{i})\pm\Omega_{22})$ to four.  The remaining four transitions involving both Rabi frequencies $\omega_{L}\pm(\Omega(\textbf{r}_{2})\pm\Omega(\textbf{r}_{1}))$ are hardly visible.

In this case, distance measurement is possible in two ways. First, the center of the sideband peak doublets correspond to $\Omega(\textbf{r}_{\mu})$, such that again a direct position determination of the two atoms is possible via Eq.~(\ref{rabifreq}). Second, the doublets are split by the dipole-dipole coupling strength $\Omega_{22}$. Thus, from Eq.~(\ref{om22}), again the distance can be obtained. Best results are obtained by a combination of the two methods.

\begin{figure}[t]
\includegraphics[width=0.9\linewidth ]{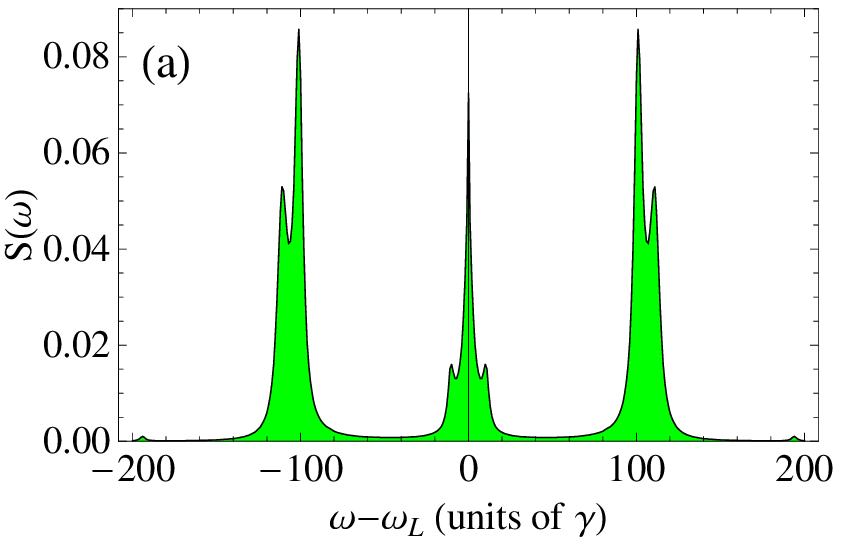}
\includegraphics[width=\linewidth ]{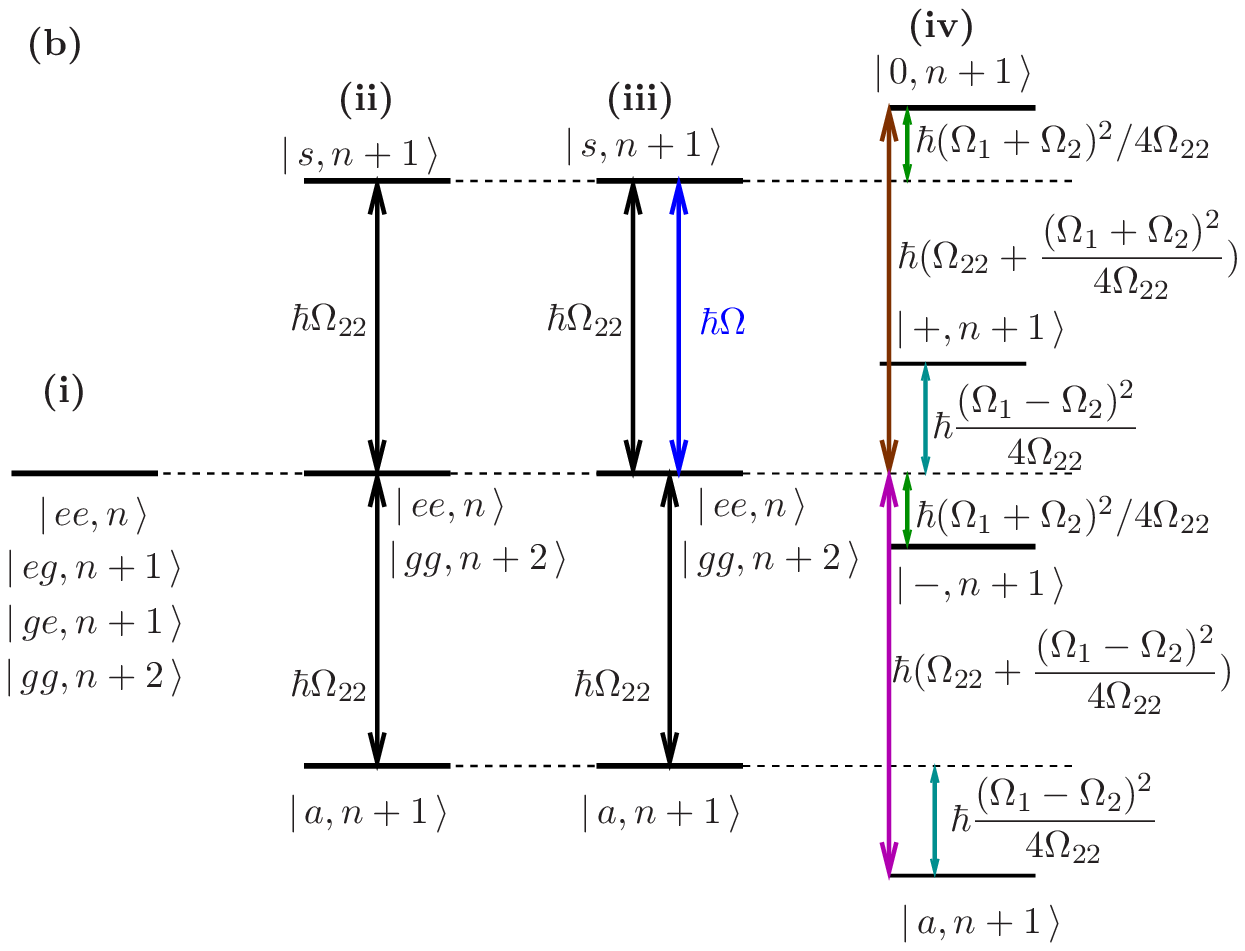}
\caption{\label{small-perturb}(Color online) (a) Incoherent spectrum of resonance fluorescence. The parameters are as in Fig.~\ref{fig-small} except for $\Omega=75\gamma$, such that the dynamics is dominated by the dipole-dipole interaction, but notably perturbed by the driving field.
(b) Dressed state representation. (i) uncoupled states. (ii) includes the dipole-dipole splitting. (iii) indicates the additional coupling to the laser field, and (iv) shows the full dressed states induced by the dipole-dipole coupling perturbed by the laser field Rabi frequencies, $\Omega_{i}$ means $\Omega(\textbf{r}_i)$.}
\end{figure}


At small interatomic separation (approx.  $R\leq\lambda/30$), the dipole-dipole interaction typically dominates the system dynamics. In Fig.~\ref{fig-small}(a), two peaks can be observed in the spectrum of resonance fluorescence which occur at the frequencies $\omega_L\pm\Omega_{22}$. These peaks reflect the well-known symmetric and anti-symmetric states $(|eg\rangle \pm |ge\rangle)/\sqrt{2}$ formed in a dipole-dipole interacting system of two two-level atoms as shown in Fig.~\ref{fig-small}(b). The two peaks correspond to the transitions $|e\rangle\to|s\rangle$ and $|s\rangle\to|g\rangle$, while the anti-symmetric state decouples from the dynamics in the limit $R\to 0$. Thus, from the positions of the peaks, one can immediately obtain the dipole-dipole interaction energy and thus the interatomic distance via Eq.~(\ref{om22}).

We finally analyze the case in which a laser field perturbs the spectrum generated due to a strong dipole-dipole interaction. In Fig.~\ref{small-perturb}(a), the driving laser field with Rabi frequency at the anti-nodes $\Omega=75\gamma$ leads to $\Omega(\textbf{r}_{1})=23.18\gamma$ and $\Omega(\textbf{r}_{2})=40.19\gamma$, which is smaller than the magnitude of relevant dipole-dipole coupling $\Omega_{22}=91.64\gamma$.
Instead of two side peaks, one now obtains split peaks at the sides as well as in the middle. 
The dressed state picture of this situation is shown in Fig.~\ref{small-perturb}(b), assuming $\Omega(\textbf{r}_{\mu})\ll\Omega_{22}$. Starting from the non-interacting system in (i), in part (ii), the dominating dipole-dipole interaction is included. It combines the states $|eg,n+1\rangle$ and $|ge,n+1\rangle$ to form $|s,n+1\rangle$ and $|a,n+1\rangle$ as symmetric and anti symmetric combinations, respectively, and shifts their energies  by the amount of the dipole-dipole interaction energy. In part (iii), the laser field is included into the dynamics. After another basis transformation to also dress the system with the laser field, this  results in the further shifting of the eigenstates by approximately $[\Omega(\textbf{r}_{1})\pm\Omega(\textbf{r}_{2})]^{2}/(4\Omega_{22})$, as shown in (iv).  The corresponding eigenenergies and eigenstates are listed in Tab.~\ref{table3}.

Sixteen transitions between the eigenkets having $n+1$ and $n$ quanta take place. Transitions $|0,n+1\rangle\rightarrow|0,n\rangle$, $|+,n+1\rangle\rightarrow|+,n\rangle$, $|-,n+1\rangle\rightarrow|-,n\rangle$ and $|a,n+1\rangle\rightarrow|a,n\rangle$ occur at transition frequency $\omega_{L}$. Transitions $|0,n+1\rangle\rightarrow|+,n\rangle$ and $|+,n+1\rangle\rightarrow|0,n\rangle$ have respective frequencies $\omega_{L}\pm\Omega_{22}\pm\Omega(\textbf{r}_{1})\Omega(\textbf{r}_{2})/\Omega_{22}$. The transitions $|0,n+1\rangle\rightarrow|-,n\rangle$ and $|-,n+1\rangle\rightarrow|0,n\rangle$ involve frequency differences equal to $\omega_{L}\pm\Omega_{22}\pm(\Omega(\textbf{r}_{1})+\Omega(\textbf{r}_{2}))^2/(2\Omega_{22})$, respectively. The separation in frequency for the transitions $|0,n+1\rangle\rightarrow|a,n\rangle$ and $|a,n+1\rangle\rightarrow|0,n\rangle$ approximates to $\omega_{L}\pm2\Omega_{22}\pm(\Omega^2(\textbf{r}_{1})+\Omega^2(\textbf{r}_{2}))/(2\Omega_{22})$, respectively. Finally the transitions $|+,n+1\rangle\rightarrow|-,n\rangle$ and $|-,n+1\rangle\rightarrow|+,n\rangle$ correspond to the frequency difference $\omega_{L}\pm(\Omega^2(\textbf{r}_{1})+\Omega^2(\textbf{r}_{2}))/(2\Omega_{22})$, respectively. These transition frequencies are the positions of the peaks in the resonance fluorescence spectrum of Fig.~\ref{small-perturb}(a). The corresponding frequencies for the transitions $|+,n+1\rangle\rightarrow|a,n\rangle$ and $|a,n+1\rangle\rightarrow|+,n\rangle$ are $\omega_{L}\pm\Omega_{22}\pm(\Omega(\textbf{r}_{1})-\Omega(\textbf{r}_{2}))^2/(2\Omega_{22})$ and for the transitions $|-,n+1\rangle\rightarrow|a,n\rangle$ and $|a,n+1\rangle\rightarrow|-,n\rangle$ are $\omega_{L}\pm\Omega_{22}\mp\Omega(\textbf{r}_{1})\Omega(\textbf{r}_{2})/\Omega_{22}$, respectively. The last four transitions do not show up in the spectrum.

\begin{table}[t]
\begin{center}
\begin{tabular}{ l l l  }
\hline \hline
  State & Composition & Energy \\
\hline 
  $|a,n+1\rangle$ & $(0,-1,1,0)/\sqrt{2}$ & $-\hbar(\Omega_{22}+(\Omega(\textbf{r}_{1})-\Omega(\textbf{r}_{2}))^2/(4\Omega_{22}))$  \\
  $|+,n+1\rangle$ & $(-1,0,0,1)/\sqrt{2}$ & $\hbar(\Omega(\textbf{r}_{1})-\Omega(\textbf{r}_{2}))^2/(4\Omega_{22})$  \\
  $|-,n+1\rangle$ & $(1,0,0,1)/\sqrt{2}$ & $-\hbar(\Omega(\textbf{r}_{1})+\Omega(\textbf{r}_{2}))^2/(4\Omega_{22})$  \\
  $|0,n+1\rangle$ & $(0,1,1,0)/\sqrt{2}$ & $\hbar(\Omega_{22}+(\Omega(\textbf{r}_{1})+\Omega(\textbf{r}_{2}))^2/(4\Omega_{22}))$ \\
\hline \hline
\end{tabular}
\end{center}

\caption{\label{table3}Eigenvectors and eigenvalues of two identical two-level atoms in the limit $\Omega(\textbf{r}_{\mu})\ll\Omega_{22}$.}
\end{table}



\subsubsection{\label{unknown}Unknown orientation of the atoms}

The methods presented so far allow for a determination of the interatomic distance if the orientation of the two atoms is known and fixed. Often, however, the orientation is unknown. Therefore, in this section, we turn to our main results, and present a method to obtain the interatomic distance for arbitrary orientations. This method can be applied if the interatomic distance is sufficiently small, such that the dipole-dipole interaction dominates the system dynamics. Fortunately, this usually is exactly the parameter range in which a distance determination is desired. In order to explain the method, we first imagine the two atoms without any driving fields. Then, it turns out that the eigenenergies of the dressed states are independent of the orientation of the two atoms~\cite{MK}. An interpretation of this fact is that without external field, there is no preferred direction in space, such that the energies cannot depend on the orientation. Since the spontaneously emitted light is emitted at frequencies corresponding to the dressed state energies, it follows that the positions of the peaks in the fluorescence spectrum of the atoms are unaffected by the orientation of the atoms. This property is approximately preserved if the atoms are driven by a weak driving field, which has a Rabi frequency much smaller than the dipole-dipole couplings. We therefore find that at small distance and weak driving, the resonance fluorescence spectrum has peak positions independent of the alignment of the two atoms. Only the relative widths and heights of the spectral features change with the orientation. It is important to note that these properties of the two-atom system are only described correctly if all dipole-dipole couplings are included in the modelling~\cite{MK}. This is the reason why we included complete Zeeman manifolds in our analysis. 

\begin{figure}
\includegraphics[width=0.9\linewidth ]{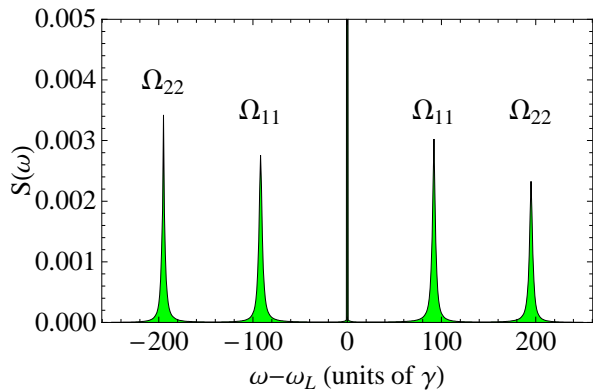}
\caption{\label{arb-orient}(Color online) Incoherent spectrum of resonance fluorescence for small distance and weak driving fields. The parameters are $R=0.04 \lambda$, $\phi=\pi/15$, $\theta=\pi/5$ and $\Omega=20\gamma$.}
\end{figure}

It remains to deduce the interatomic distance from the peak positions in the resonance fluorescence spectrum. For this, we again analyze the eigenvalues of the interaction Hamiltonian, which determine the peak positions. Since the peak positions and thus these eigenvalues are independent of the orientation, it suffices to  evaluate analytic expressions for the peak positions in a simple configuration. Investigating the eigenvalues for $\theta=0$, and assuming $\Omega(\textbf{\textit{r}}_{\mu})\ll\Omega_{ii}$, we find that the eigenenergies are given by $0$, $\pm\Omega_{11}(\theta=0)$, and $\pm \Omega_{22}(\theta=0)$.
An example for this is shown in Fig.~\ref{arb-orient}. The four side peaks are located at $\pm\Omega_{11}$ and $\pm\Omega_{22}$, and the interatomic distance can be gained from both coupling constants via Eq.~(\ref{om11}) and (\ref{om22}). As can be seen from Fig.~\ref{fig-coupling}, for $\theta\in\{0,\pi\}$, the coupling constant $|\Omega_{22}|$ is larger than $|\Omega_{11}|$ for small interatomic distances $R$. Thus, the inner [outer] peaks in Fig.~\ref{arb-orient} correspond to 
$|\Omega_{11}|$ [$|\Omega_{22}|$]. Due to the dependence of the amplitude of the spectral peaks on the orientation, the peaks at $\pm\Omega_{11}$ visible in Fig.~\ref{arb-orient} may be suppressed. For example, in Fig.~\ref{fig-small}(a), we found only a single pair of sidebands corresponding to $\pm\Omega_{22}$. This can be understood by observing that these peaks at $\pm\Omega_{22}$ correspond to states $|2\rangle$ populated by the driving laser field, while states $|1\rangle$ and $|3\rangle$ are only populated in certain geometries.

We end this section by noting that in the case of large interatomic separation, the dipole-dipole interaction vanishes, such that the interatomic coupling also becomes independent of the orientation. The driving field Rabi frequencies experienced by the two atoms, however, may not be the same, as they depend on the scalar product $\bf{k}_L\cdot \bf{r}_\mu$.  Thus, by applying a driving field with larger Rabi frequency, it is possible to measure the position of the atoms projected on the propagation axis of the driving field. For arbitrary orientations, however, a single measurement of this type does not allow to deduce the interatomic distance since the position transverse to the wave vector remains unknown.

We thus conclude that by applying weak driving fields, the interatomic distance can be measured from a pair of nearby atoms independent of their mutual orientation, as long as the dipole-dipole interaction is strong enough to dominate the system dynamics.

\begin{figure}
\includegraphics*[width=0.8\linewidth,angle=0]{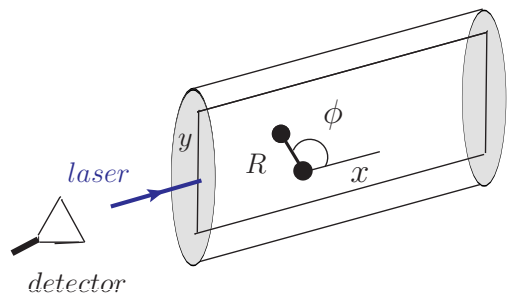}
\caption{\label{fig-2d}(Color online) Setup with the two nearby atoms confined inside a two-dimensional waveguide. The atoms are located in the $x$-$y$ plane. }
\end{figure} 

\subsection{\label{orientation}Determination of the orientation}

So far, we have discussed techniques for the measurement of the interatomic distance, and have demonstrated how the distance can be measured independent of the orientation of the interparticle distance vector. In this Section, we augment our analysis by discussing the determination of the relative orientation of the two atoms. We discuss two different cases, corresponding to two different methods to determine the orientation. First, we discuss the case of unknown $\phi$, assuming $\theta=\pi/2$. This case corresponds to an effective two-dimensional geometry of the system which can be realized, e.g., by embedding the atoms in a planar matter waveguide~\cite{waveguide1, waveguide2, waveguide3}. In this case, the orientation is deduced from the $\phi$-dependent peak positions in the fluorescence spectrum induced by the driving laser field. Second, we study the case of unknown $\theta$  and $\phi=\pi/2$. This corresponds to atoms on a surface, driven by a laser field propagating perpendicular to the surface. In this case, we will determine $\theta$ via the resonance fluorescence intensity emitted in a particular direction.

\begin{figure}
\includegraphics*[width=0.9\linewidth,angle=0]{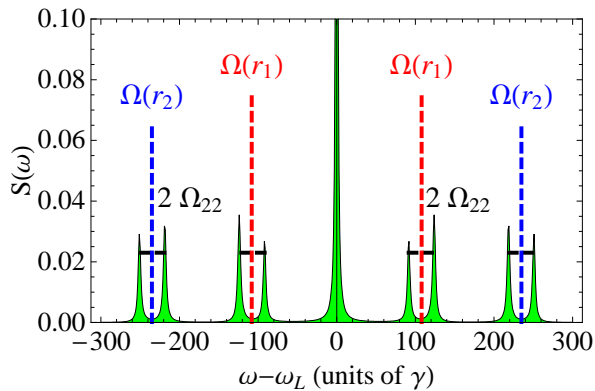}
\caption{\label{2d-strong}(Color online) Incoherent spectrum of resonance fluorescence for two atoms in a geometry as shown in Fig.~\ref{fig-2d}. The parameters are $R=0.07\lambda$, $\theta=\pi/2$, $\phi=0.1\pi$, and $\Omega=350\gamma$.}
\end{figure} 

\subsubsection{\label{unknown-phi}Unknown $\phi$: Planar waveguide}

In this section, we assume that the two atoms are confined in the $x$-$y$ plane ($\theta=\pi/2$) as shown in Fig.~\ref{fig-2d}, as it is the case, for example, in a planar waveguide. Since such a setup may also constrain the observation direction, we assume detection in a direction anti-parallel to the incident driving field, i.e., along the $-x$ direction. This way, the spectrum can be measured without background from the incident laser field.
In this geometry, the coupling constants $\Omega_{21}$ and $\Omega_{32}$ are zero for all values of $\phi$. Since only the second transition $|2\rangle\leftrightarrow|4\rangle$ is driven, the populations of the levels $|1\rangle$ and $|3\rangle$ are zero for the whole range of $\phi$. 
The parallel coupling constants $\Omega_{11}=\Omega_{33}$ and $\Omega_{22}$ are independent of $\phi$. Nevertheless, the spectra depend strongly on $\phi$ because of the $\phi$-dependence of the Rabi frequency $\Omega(\textbf{r}_{2})$. 
The obtained spectra are identical if $\phi$ is replaced by $2\pi-\phi$.

From the results of Sec.~\ref{unknown} it is clear that the peak positions in the resonance fluorescence spectrum cannot be used to determine the orientation as long as the dipole-dipole interaction dominates the dynamics. Therefore, we apply stronger driving fields, such that the external driving dominates the dynamics.  

In this case, for most values of $\phi$, a typical spectrum obtained is shown in Fig.~\ref{2d-strong}. Using the results of Sec.~\ref{fixed}, we conclude that the two doublets on each side corresponding to Mollow sidebands at the two Rabi frequencies experienced by the atoms, split by the dipole-dipole interaction. The doublets can thus be used to approximately read off the two position-dependent Rabi frequencies. Assuming that the distance is known from a measurement with a weaker driving field as described in Sec.~\ref{unknown}, the components of the position vectors of the individual atoms as well as the relative alignment of the atoms along the laser can be found out by using the position dependent Rabi frequencies $\Omega(\textbf{r}_{\mu})$ as follows:
\begin{equation} \label{eq-phi}
\phi=\cos^{-1} \left\{\dfrac{1}{\textit{k}_{L}R} \left [ \sin^{-1}\left (\dfrac{\Omega(\textbf{r}_{2})}{\Omega} \right)-\sin^{-1}\left(\dfrac{\Omega(\textbf{r}_{1})}{\Omega}\right)\right]\right\}\,.
\end{equation}

In Fig.~\ref{2d-strong}, the peaks in the spectrum of resonance fluorescence occur at approximately $0$, $\pm91.68\gamma$, $\pm123.95\gamma$, $\pm218.59\gamma$, and  $\pm250.86\gamma$. Considering the mean values of the peak separation in the inner side band doublets as $\Omega(\textbf{r}_{1})$ and that in the exterior side band doublets as $\Omega(\textbf{r}_{2})$, from Eq.~(\ref{eq-phi}) we obtain $0.091\pi$ as the value of $\phi$, which deviates from the true value by about $9\%$. This deviation can be attributed to the imperfect determination of the Rabi frequencies as the mean value of the two peaks in the doublets.

Increasing $\phi$ from the value that has been used in Fig.~\ref{2d-strong}, the position-dependent Rabi frequencies change, until the sideband doublets corresponding to the two Rabi frequencies start to overlap, as the two position-dependent Rabi frequencies approach each other. In this case, it is difficult to estimate $\phi$ directly from the spectrum, since a clear identification of the different peaks is not obvious. One strategy is to increase the driving field intensity. Since the peak separation $|\Omega({\bf r}_1) - \Omega({\bf r}_2)|$ is proportional to $\Omega$, this increase eventually leads to a splitting larger than the line widths of the involved peaks, such that an identification becomes possible. In any case, it can be concluded from overlapping peaks that $\phi$ is close to $\pi/2$ or $3\pi/2$, since then $\Omega(\textbf{r}_{1})\approx\Omega(\textbf{r}_{2})$.

In summary, the relative orientation of the two atoms with respect to the laser can be determined using stronger laser fields. This works well if the position dependent Rabi frequencies are different from each other, since then the spectral lines are well separated. This is the case for $\phi$ not close to $\pi/2$ or $3\pi/2$. Accordingly, if the corresponding spectral peaks overlap, it can be concluded that $\phi$ is close to $\pi/2$ or $3\pi/2$.

\subsubsection{\label{unknown-theta}Unknown $\theta$: Atoms on a surface}

\begin{figure}
\includegraphics*[width=0.9\linewidth,angle=0]{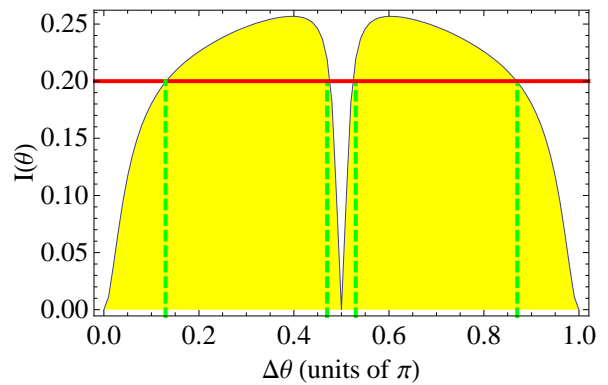}
\caption{\label{int-theta}(Color online) Fluorescence intensity emitted on the $\sigma$-transitions observed in $z$ direction from two atoms on a surface. $\Delta \theta$ is the relative angle between laser polarization and interatomic distance vector. The green dashed lines indicate the four possible values of $\Delta \theta$ corresponding to a possible measured intensity indicated by the solid red line. The parameters are $R=0.07 \lambda$, $\phi=\pi/2$ and $\Omega=200\gamma$.}
\end{figure} 

In this section, we consider the case of two atoms on a surface, driven by a laser field propagating perpendicular to the surface. Thus, $\phi$ is fixed to $\pi/2$, while $\theta$ is unknown. In this case, the two Rabi frequencies experienced by the atoms are equal and independent of $R$ and $\theta$. Therefore, in contrast to the previous Sec.~\ref{unknown-phi}, here, we determine the value of $\theta$ with the help of the resonance fluorescence intensity. In particular, we consider the $\sigma$-intensity emitted by the dipoles $\textbf{\textit{d}}_{1}$ and $\textbf{\textit{d}}_{3}$, measured by a detector placed in $z$ direction since there is no $\sigma$-spectrum in $y$ direction.

It turns out that the configuration is symmetric in the sense that a rotation of the laser polarization and the detectors around the $x$ direction is equivalent to a corresponding rotation of the interatomic distance vector. Therefore, the measured resonance fluorescence intensity depends only on the relative angle $\Delta \theta$ between the laser polarization direction and the orientation of the two atoms $\theta$ on the surface. 
In Fig.~\ref{int-theta}, we show this resonance fluorescence intensity  versus the relative angle $\Delta \theta$. A plot like this can either be recorded by rotating the sample in the $y$-$z$ plane, or by rotating the polarization vector of the laser field around its propagation axis.  

From Fig.~\ref{int-theta}, we find that the intensity is symmetric around $\Delta \theta=\pi/2$, and it is easy to see that one value of $\sigma$ intensity corresponds to at  most four values of $\Delta \theta$. These four values of $\Delta  \theta$ can be roughly divided into the four ranges $0\rightarrow\pi/4$,  $\pi/4\rightarrow\pi/2$, $\pi/2\rightarrow3\pi/4$ and $3\pi/4\rightarrow\pi$, respectively. This can be understood by noting that the orthogonal coupling constants responsible for the $\sigma$ intensity are the same for orientations $\theta$, $\pi/2-\theta$, $\pi/2+\theta$ and $\pi-\theta$. At $\Delta \theta\in\{0, \pi/2, \pi\}$, the $\sigma$ intensity is zero since the orthogonal dipole dipole coupling constants $\Omega_{32}$ and $\Omega_{21}$ vanish at these points, see Fig.~\ref{fig-coupling}(a). Thus there is no population in states $|1\rangle$ and $|3\rangle$, and the intensity of light emitted from these states is zero.
The points of zero $\sigma$ intensity $\Delta \theta\in\{0,\pi/2,\pi\}$ correspond to situations in which the polarization vector is parallel, perpendicular, or anti-parallel to $\textbf{\textit{R}}$, respectively.
Since these values of $\Delta \theta$ with vanishing intensity can easily be identified, they allow to determine $\theta$ from the amount of sample or driving field polarization rotation required to reach these values. In particular, the symmetry point $\Delta \theta=\pi/2$  is well-suited for such a measurement.

\section{Summary and Discussion}

We have discussed methods to measure the relative distance and orientation of two nearby atoms in arbitrary geometry. Our methods are based on the driving of the two atoms with a standing wave field, and on detection of the resonance fluorescence intensity and spectrum in the far field. The distance and orientation information is encoded in the scattered light via the position-dependent Rabi frequencies and via the distance- and orientation-dependent dipole-dipole couplings. Since unlike in previous studies, we consider the case of arbitrary orientation, the atoms must be described using complete Zeeman manifolds in order to correctly model all relevant dipole-dipole couplings between parallel dipole moments as well as between orthogonal ones. 

As preliminary work, we have analyzed the fluorescence spectra in particular known geometries, in order to identify dressed-state interpretations in the various limiting cases of interest. These in particular are the case of dominating laser-induced dynamics perturbed by the dipole-dipole interaction, and the case of dominating dipole-dipole coupling modified by the presence of a weaker laser field.
Next, we have shown that the case of dominating dipole-dipole interaction enables one to measure the distance between two nearby particles independent of the relative orientation. The reason for this is that the eigenvalues of the total Hamiltonian describing the dynamics, and thus the position of the system dressed states, are independent of the orientation if the two atoms are undriven. We found that a weak driving field allows to probe these dressed states without perturbing the independence on the orientation. Finally, we discussed the measurement of the relative orientation. We presented two methods. The first is based on the position-dependent Rabi frequencies, which under certain conditions reveal the orientation of the two particles. The second method is based on the measurement of the resonance fluorescence intensity in a particular direction. This intensity is a measure for the population in the excited states not driven by the laser field, and therefore a signature for the magnitude of the dipole-dipole coupling between orthogonal dipole moments. We applied the two methods to the two cases of atoms confined in a two-dimensional waveguide, and to atoms on a surface, in which either the polar or the azimuthal angle of the interatomic distance vector is known. 

In principle, these methods to determine the orientation can also be applied for the determination of both polar and azimuthal angle. The most promising ansatz is to make use of the resonance fluorescence intensity in a particular direction as discussed in Sec.~\ref{unknown-theta} together with a rotation of the sample or of the driving laser polarization in order to fix one of the two angles at a value which renders the spectrum simpler (e.g., $0$, $\pi/2$ or $\pi$). Then, the methods described in Secs.~\ref{unknown-phi} and \ref{unknown-theta}
can be used to determine the other angle. The most straightforward implementation, however, strongly depends on the experimental possibilities to modify the setup. For example, in many cases, a rotation of the sample will be difficult. 

For many applications, the generalization to more than two particles is desirable. It remains to be seen whether methods based on the dipole-dipole interaction can also be applied in such cases. One approach could be to combine methods presented here together with a selective addressing of individual atoms at least in one or two dimensions, for example, by position-dependent state transfer.

\acknowledgments

QG acknowledges support by the Higher Education Commission (HEC), Pakistan, the International Max Planck Research School for Quantum Dynamics in Physics, Chemistry and Biology, Heidelberg, Germany, and by Deutscher Akademischer Austauschdienst (DAAD).


\end{document}